\documentclass[a4paper,twoside]{article}
\usepackage{graphicx}
\usepackage{jcmse}

\newcommand{\QED}{\hspace*{\fill}\rule{2.5mm}{2.5mm}}

\newcommand{\JCMSEtitle}%
{Multifractal Analysis of Various Probability Density Functions
in Turbulence}

\newcommand{\JCMSEauthor}
{\textbf{Toshihico Arimitsu\footnote{arimitsu@cm.ph.tsukuba.ac.jp}\footnote{Corresponding 
author}}\\[1ex]
Institute of Physics, University of Tsukuba,
Ibaraki 305-8571, Japan\\[1ex]
\textbf{Naoko Arimitsu\footnote{arimitsu@ynu.ac.jp}}\\[1ex]Graduate School of EIS, Yokohama Nat'l.~University, 
Yokohama 240-8501, Japan}

\newcommand{\be}{\begin{equation}}
\newcommand{\ee}{\end{equation}}
\newcommand{\bea}{\begin{eqnarray}}
\newcommand{\eea}{\end{eqnarray}}

\newcommand{\aeq}{&=&}

\newcommand{\itDelta}{{\it \Delta}}

\newcommand{\itPi}{{\it \Pi}}

\newcommand{\bra}{\langle}
\newcommand{\ket}{\rangle}
\newcommand{\dbra}{\bra \! \bra}
\newcommand{\dket}{\ket \! \ket}

\newcommand{\bq}{{\bar q}}

\newcommand{\rRe}{{\rm Re}}

\newcommand{\rS}{{\rm S}}

\begin{document}

\maketitle
\markboth{\JCMSEauthorshort{T.~Arimitsu and N.~Arimitsu}}
     {\JCMSEtitleshort{Multifractal Analysis of Various PDF's}}

\centerline{Received 27 January, 2003}

\begin{abstract}
The probability density functions measured 
by Lewis and Swinney for turbulent Couette-Taylor flow, 
observed by Bodenschatz and co-workers 
in the Lagrangian measurement of particle accelerations
and those obtained in the DNS by Gotoh et al.\
are analyzed in excellent agreement with the theoretical formulae 
derived with the multifractal analysis,
a unified self-consistent approach based on 
generalized entropy, i.e., the Tsallis or the Renyi entropy. 
This analysis rests on the invariance of the Navier-Stokes equation 
under a scale transformation for high Reynolds number, 
and on the assumption that the distribution of the exponent 
$\alpha$, introduced in the scale transformation, is multifractal and 
that its distribution function is given by taking extremum of 
the generalized entropy with the appropriate constraints. 
It also provides analytical formula for the scaling exponents of 
the velocity structure function which explains 
quite well the measured quantities in experiments and DNS.
\end{abstract}



\begin{MSC}
Mathematical modeling of turbulent processes and determinated chaos
\end{MSC}

\section{Introduction}
\label{intro}

The {\it multifractal analysis} of turbulence 
\cite{AA}-\cite{AA10} 
is a unified self-consistent approach for the systems with large deviations,
which has been constructed based on 
the Tsallis-type distribution function \cite{Tsallis88,Tsallis99}
that provides an extremum of the {\it extensive} R\'{e}ny \cite{Renyi} 
or the {\it non-extensive} Tsallis entropy \cite{Tsallis88,Tsallis99,Havrda-Charvat}
under appropriate constraints.
The analysis rests
on the scale invariance of the Navier-Stokes equation for high Reynolds number, 
and on the assumptions that the singularities due to the invariance 
distribute themselves multifractally in physical space.
The multifractal analysis is a generalization of 
the log-normal model \cite{Oboukhov62}-\cite{Yaglom}. It has been shown \cite{AA4} that
the multifractal analysis derives the log-normal model 
when one starts with the Boltzmann-Gibbs entropy.

In this paper, we derive the formula for various probability density functions (PDF's) 
in fully developed turbulence by means of 
the multifractal analysis, and analyze the PDF's observed in three experiments. 
The first is the PDF of velocity fluctuations at $R_\lambda = 262$ of 
the Taylor microscale Reynolds number measured by Lewis and Swinney \cite{Lewis-Swinney99} 
in the real experiment for turbulent Couette-Taylor flow 
in a concentric cylinder system.
The second is the PDF of accelerations at $R_\lambda = 970$ obtained in 
the Lagrangian measurement of particle accelerations 
that was realized by Bodenschatz and co-workers \cite{EB01a,EB01b} 
by raising dramatically the spatial and temporal measurement resolutions
with the help of the silicon strip detectors.
The third is the PDF's of velocity fluctuations,
of velocity derivatives and of fluid particle accelerations 
at $R_\lambda = 380$ that was extracted by Gotoh et al.\ from
the DNS of the size 1024$^3$ \cite{Gotoh02}.

For high Reynolds number $\rRe \gg 1$, or for the situation where 
effects of the kinematic viscosity $\nu$ can be neglected compared with
those of the turbulent viscosity, the Navier-Stokes equation,
$
\partial {\vec u}/\partial t
+ ( {\vec u}\cdot {\vec \nabla} ) {\vec u} 
= - {\vec \nabla} \left(p/\rho \right)
+ \nu \nabla^2 {\vec u}
\label{N-S eq}
$,
of an incompressible fluid is invariant under 
the scale transformation~\cite{Frisch-Parisi83,Meneveau87b}
$
{\vec r} \rightarrow \lambda {\vec r}
$, 
$
{\vec u} \rightarrow \lambda^{\alpha/3} {\vec u}
$, 
$
t \rightarrow \lambda^{1- \alpha/3} t
$ and
$
\left(p/\rho\right) \rightarrow \lambda^{2\alpha/3} \left(p/\rho\right)
$
where the exponent $\alpha$ is an arbitrary real quantity.
The quantities $\rho$ and $p$ represent, respectively, mass density and pressure.
The Reynolds number $\rRe$ of the system is given by 
$
{\rm Re} = \delta u_{\rm in} \ell_{\rm in}/\nu = ( \ell_{\rm in}/\eta )^{4/3}
$
with the Kolmogorov scale 
$
\eta = ( \nu^3/\epsilon )^{1/4}
$~\cite{K41}
where $\epsilon$ is the energy input rate at the input scale 
$\ell_{\rm in}$.
Here, we introduced 
$\delta u_{\rm in} = \vert u(\bullet + \ell_{\rm in}) - u(\bullet) \vert
$ 
with the definition of the velocity fluctuation 
$
\delta u_n = \vert u(\bullet + \ell_n) - u(\bullet) \vert
$
where $u$ is a component of velocity field $\vec{u}$, and
$\ell_n$ is a distance between two points.
The {\it pressure} (divided by the mass density) difference 
$
\delta p_n = \vert p/\rho(\bullet + \ell_n) - p/\rho(\bullet) \vert
$
between two points separated by the distance $\ell_n$ is another
important observable quantity.
We are measuring distance by the discrete units 
$
\ell_n = \delta_n \ell_0
\label{r-n}
$
with $\delta_n = 2^{-n}$ $(n=0,1,2,\cdots)$.
The non-negative integer $n$ can be interpreted as 
the {\it multifractal depth}.\footnote
{
The multifractal depth $n$ is related to 
the number of steps within the energy cascade model .
At each step of the cascade, say at the $n$th step, eddies break up into 
two pieces producing an energy cascade with the energy-transfer rate
$\epsilon_n$ that represents the rate of transfer of energy per unit mass 
from eddies of size $\ell_n$ to those of size $\ell_{n+1}$.
The energy dissipation rate becomes singular in the limit $n \rightarrow \infty$
for $\alpha < 1$,
i.e.,
$
\lim_{n \rightarrow \infty} \epsilon_n / \epsilon_0 
= \lim_{n \rightarrow \infty} \left( \ell_n/\ell_0 \right)^{\alpha -1}
\rightarrow \infty
$.
}
However, we will treat it as positive real number in the analysis of experiments.

Let us consider the quantity 
$
\delta x_n = \vert x(\bullet + \ell_n) - x(\bullet) \vert
$
having the scaling property
$
\vert x_n \vert \equiv \vert \delta x_n / \delta x_0 \vert
= \delta_n^{\kappa \alpha/3}
$.
Its spatial derivative is defined by
$
\vert x' \vert = \lim_{\ell_n \rightarrow 0} \delta x_n /\ell_n 
\propto \lim_{n \rightarrow \infty} \ell^{\kappa \alpha /3 -1}
$
which becomes singular for 
$
\alpha < 3/\kappa
$.
The values of exponent $\alpha$ specify the degree of singularity.
We see that the scale invariance provides us with
$
\delta u_n /\delta u_0 = \delta_n^{\alpha/3}
$
and
$
\delta p_n / \delta p_0 = (\ell_n / \ell_0)^{2\alpha/3}
\label{p-alpha}
$
giving, respectively,
$
\kappa = 1
$
for the velocity fluctuation and
$
\kappa = 2
$
for the pressure fluctuation.
The velocity derivative and 
the fluid particle acceleration may be estimated, respectively, by 
$
\vert u^\prime \vert = \lim_{n \rightarrow \infty} u^\prime_n
$
and by
$
\vert \vec{\mathrm{a}} \vert = \lim_{n \rightarrow \infty} \mathrm{a}_n
$
where we introduced the velocity derivative
$
u'_n = \delta u_n / \ell_n
$
and the acceleration 
$
\mathrm{a}_n = \delta p_n / \ell_n
$
corresponding to the characteristic length $\ell_n$.
Note that the acceleration $\vec {\mathrm{a}}$ of a fluid particle is given by
the substantive time derivative of the velocity:
$
{\vec {\mathrm a}} = \partial {\vec u}/\partial t
+ ( {\vec u}\cdot {\vec \nabla} ) {\vec u}
$.
We see that the velocity derivative and 
the fluid particle acceleration become singular for $\alpha < 3$ and $\alpha < 1.5$, 
respectively,
i.e.,
$
\vert u' \vert \propto \lim_{\ell_n \rightarrow 0} \ell_n^{(\alpha/3)-1}
\rightarrow \infty
$
and
$
\vert \vec{a} \vert \propto \lim_{\ell_n \rightarrow 0} \ell_n^{(2\alpha/3)-1}
\rightarrow \infty
$.

\section{Multifractal analysis}

The multifractal analysis rests on the multifractal distribution of $\alpha$.
The probability 
$
P^{(n)}(\alpha) d\alpha
$
to find, at a point in physical space, a singularity labeled by an exponent 
in the range 
$
\alpha \sim \alpha + d \alpha
$
is given by~\cite{AA1}-\cite{AA4}
$
P^{(n)}(\alpha) = \left[ 1 - (\alpha - \alpha_0)^2 \big/ (\itDelta \alpha )^2 
\right]^{n/(1-q)}/Z_{\alpha}^{(n)} 
\label{Tsallis prob density}
$
with an appropriate partition function
$
Z_{\alpha}^{(n)}
$
and
$
(\itDelta \alpha)^2 = 2X \big/ [(1-q) \ln 2 ]
$.
The range of $\alpha$ is $\alpha_{\rm min} \leq \alpha \leq \alpha_{\rm max}$ with
$
\alpha_{\rm min} = \alpha_0 - \itDelta \alpha
$, 
$
\alpha_{\rm max} = \alpha_0 + \itDelta \alpha
$.
This is consistent with the relation \cite{Meneveau87b,AA4}
$
P^{(n)}(\alpha) \propto \delta_n^{1-f(\alpha)}
$
that is a manifestation of scale invariance and 
reveals how densely each singularity, labeled by $\alpha$, 
fills physical space.
In the present model, the multifractal spectrum $f(\alpha)$
is given by~\cite{AA1}-\cite{AA4}
$
f(\alpha) = 1 + (1-q)^{-1} \log_2 [ 1 - (\alpha - \alpha_0 )^2
/ (\Delta \alpha )^2 ]
\label{Tsallis f-alpha}
$.
In spite of the different characteristics of the entropies,
i.e., extensive and non-extensive,
the distribution functions $P^{(n)}(\alpha)$ giving their extremum
have the common structure.~\footnote{
Within the present formulation, the decision cannot be pronounced 
which of the entropies is underlying the system of turbulence.
}

The dependence of the parameters $\alpha_0$, $X$ and $q$ on 
the intermittency exponent $\mu$ is determined, 
self-consistently, with the help of the three independent equations, i.e.,
the energy conservation:
$
\left\bra \epsilon_n \right\ket = \epsilon
\label{cons of energy}
$,
the definition of the intermittency exponent $\mu$:
$
\bra \epsilon_n^2 \ket 
= \epsilon^2 \delta_n^{-\mu}
\label{def of mu}
$,
and the scaling relation~\footnote{
The scaling relation is a generalization of the one derived first in
\cite{Costa,Lyra98} to the case where the multifractal spectrum
has negative values.
}:
$
1/(1-q) = 1/\alpha_- - 1/\alpha_+
\label{scaling relation}
$
with $\alpha_\pm$ satisfying $f(\alpha_\pm) =0$. 
The average $\bra \cdots \ket$ is taken with $P^{(n)}(\alpha)$.

It has been shown that the probability 
$\itPi^{(n)}(x_n) dx_n$ to find a physical quantity
$
x_n
$
in the range $x_n \sim x_n+dx_n$ is given in the form
\be
\itPi^{(n)}(x_n) dx_n = \itPi^{(n)}_{\rS}(x_n) dx_n 
+ \Delta \itPi^{(n)}(x_n) dx_n
\label{def of Lambda}
\ee
with the normalization
$
\int_{-\infty}^{\infty} dx_n  \itPi^{(n)}(x_n) =1
$.
The first term represents the contribution by the singular part of the quantity
$ x_n $ stemmed from the multifractal distribution of its singularities 
in physical space. This is given by
$
\itPi^{(n)}_{\rS}(\vert x_n \vert) dx_n \propto P^{(n)}(\alpha) d \alpha
\label{singular portion}
$
with the transformation of the variables,
$
\vert x_n \vert = \delta_n^{\kappa \alpha/3}
$.
Whereas the second term $\Delta \itPi^{(n)}(x_n) dx_n$ represents 
the contribution from the dissipative term
in the Navier-Stokes equation, and/or the one 
from the errors in measurements.
The dissipative term has been discarded in the above investigation for 
the distribution of singularities since it violates 
the invariance under the scale transformation.
The contribution of the second term provides a correction to the first one.
Note that each term in (\ref{def of Lambda}) is a multiple of two 
probability functions, i.e., the one to
determine the portion of the contribution among two independent 
origins, and the other to find $x_n$ in the range $x_n \sim x_n+dx_n$.
Note also that the values of $x_n$ originated in the singularity
are rather large representing intermittent large deviations, and that those 
contributing to the correction terms are small in comparison with
its deviation.

The $m$th moment of the variable $\vert x_n \vert$ is given by 
$
\dbra \vert x_n \vert^m \dket \equiv \int_{-\infty}^{\infty} dx_n  
\vert x_n \vert^m \itPi^{(n)}(x_n) 
=2 \gamma^{(n)}_m
+ (1-2\gamma^{(n)}_0 ) \
a_{\kappa m} \ \delta_n^{\zeta_{\kappa m}}
\label{structure func m}
$
where
$
2\gamma^{(n)}_m = \int_{-\infty}^{\infty} dx_n\ 
\vert x_n \vert^m \Delta \itPi^{(n)}(x_n)
$,
$
a_{3\bq} = \{ 2 / [\sqrt{C_{\bq}} ( 1+ \sqrt{C_{\bq}} ) ] \}^{1/2}
$
with
$
{C}_{\bq} = 1 + 2 \bq^2 (1-q) X \ln 2
\label{cal D}
$.
The quantity \cite{AA1}-\cite{AA4}
\be
\zeta_m = \alpha_0 m/3 
- 2Xm^2/[9 (1+C_{m/3}^{1/2} )]
- [1-\log_2 (1+C_{m/3}^{1/2} ) ] /(1-q)
\label{zeta}
\ee
is the scaling exponent of the $m$th moment of the velocity structure function
$
\dbra \vert \delta u_n / \delta u_0 \vert^m \dket 
$,
i.e., the velocity fluctuations ($\kappa = 1$).
It explains the experimental results, successfully.
The formula (\ref{zeta}) is independent of $n$,
which is a manifestation of the scale invariance.

We now derive the PDF, $\hat{\itPi}^{(n)}(\xi_n)$, 
defined by the relation 
$
\hat{\itPi}^{(n)}(\xi_n) d\xi_n
= \itPi^{(n)}(x_n) d x_n
$
with the variable $\xi_n = x_n / \dbra x_n^2 \dket^{1/2}$ scaled by 
the deviation $\dbra x_n^2 \dket^{1/2}$. This PDF is to be compared with
the observed PDF's. The variable is related with $\alpha$ by
$
\vert \xi_n \vert = \bar{\xi}_n \delta_n^{\kappa \alpha /3 -\zeta_{2\kappa}/2}
$
with
$
\bar{\xi}_n = [2 \gamma_2^{(n)} \delta_n^{-\zeta_{2\kappa}} + (1-2\gamma_0^{(n)} ) 
a_{2\kappa}]^{-1/2}
$.
It is reasonable to imagine that the origin of intermittent rare events is 
attributed to the first singular term in (\ref{def of Lambda}), and that
the contribution from the second term is negligible. We then have, for
$\xi_n^* \leq \vert \xi_n \vert \leq \xi_n^{\rm max}$,
\bea
\hat{\itPi}^{(n)}(\xi_n) d \xi_n 
= \itPi^{(n)}_{\rm S} (x_n) dx_n
= \tilde{\itPi}^{(n)}_{\rS} \frac{\bar{\xi}_n}{\vert \xi_n \vert}
\left[1 - \frac{1-q}{n}\ 
\frac{\left(3 \ln \vert \xi_n / \xi_{n,0} \vert\right)^2}{
2\kappa^2 X \vert \ln \delta_n \vert} \right]^{n/(1-q)} d \xi_n
\label{PDF kappa large}
\eea
with
$
\xi_{n,0} = \bar{\xi}_n \delta_n^{\kappa \alpha_0 /3 -\zeta_{2\kappa} /2}
$,
$
\xi_n^{\rm max} = \bar{\xi}_n \delta_n^{\kappa \alpha_{\rm min}/3 -\zeta_{2\kappa} /2}
$,
$
\tilde{\itPi}^{(n)}_{\rS} = 3 (1-2\gamma^{(n)}_0)
/ (2\kappa \bar{\xi}_n \sqrt{2\pi X \vert \ln \delta_n \vert} )
$.
On the other hand, for smaller values, 
the contribution to the PDF comes, mainly, from thermal fluctuations 
or measurement error. It may be described by a Gaussian function, i.e.,
for
$
\vert \xi_n \vert \leq \xi_n^*
$,
{\small
\bea
\hat{\itPi}^{(n)}(\xi_n) d \xi_n =
\left[ \hat{\itPi}^{(n)}_{\rS}(x_n)
+\Delta \hat{\itPi}^{(n)}(x_n) \right] d x_n
= \tilde{\itPi}^{(n)}_{\rS}
\exp\left\{-\frac{1}{2}\left[1+\frac{3}{\kappa} f'(\alpha^*)\right] 
\left[\left(\xi_n/\xi_n^* \right)^2 -1 \right]\right\} d \xi_n.
\label{PDF kappa small}
\eea
}This specific form of the Gaussian function is determined by the condition that 
the two PDF's (\ref{PDF kappa large}) and (\ref{PDF kappa small}) 
should have the same value and the same slope at
$\xi_n^*$ which is defined by
$
\xi_n^* = \bar{\xi}_n \delta_n^{\kappa \alpha^* /3 -\zeta_{2\kappa} /2}
$
with $\alpha^*$ being the smaller solution of 
$
\zeta_{2\kappa}/2 -\kappa \alpha/3 +1 -f(\alpha) = 0
$.
It is the point at which 
$
\hat{\itPi}^{(n)}(\xi_n^*)
$
has the least $n$-dependence for large $n$.

With the help of the second equality in (\ref{PDF kappa small}),
we obtain $\Delta \itPi^{(n)}(x_n)$, and have the formula 
to evaluate $\gamma_m^{(n)}$ in the form
$
2\gamma_m^{(n)} = \left(K_m^{(n)} - L_m^{(n)}\right) \Big/
\left(1 + K_0^{(n)} - L_0^{(n)}\right)
$
where
\bea
K_m^{(n)} \aeq \frac{3\ \delta_n^{\kappa (m+1)\alpha^*/3 -\zeta_{2\kappa}/2}}
{\kappa \sqrt{2 \pi X \vert \ln \delta_n \vert}}
\int_0^1 dz\ z^{m} \exp\left\{-\frac{1}{2}
\left[1+\frac{3}{\kappa}f'(\alpha^*)\right]\left(z^2-1\right)\right\},\\
L_m^{(n)} \aeq \frac{3\ \delta_n^{\kappa m \alpha^*/3}}
{\kappa \sqrt{2 \pi X \vert \ln \delta_n \vert}}
\int_{z_{\rm min}^*}^1 dz\ z^{m-1} 
\left[1 - \frac{1-q}{n}\ \frac{\left(3 \ln \vert z / z_0^* \vert \right)^2}{
2\kappa^2 X \vert \ln \delta_n \vert} \right]^{n/(1-q)}
\eea
with
$
z_{\rm min}^* = \xi_{\rm min}/\xi_n^* 
=\delta_n^{\kappa (\alpha_{\rm max} - \alpha^*)/3}
$,
$z_0^* = \xi_{n,0}/\xi_n^* 
=\delta_n^{\kappa (\alpha_0 - \alpha^*)/3}
$.
Now, the PDF for the variable scaled by its own deviation, given by 
(\ref{PDF kappa large}) and (\ref{PDF kappa small}), is completely determined by 
the intermittency exponent $\mu$ and the multifractal depth $n$
which gives a length scale $\ell_n$.
The PDF's for velocity fluctuations and for derivatives are given by 
(\ref{PDF kappa large}) and (\ref{PDF kappa small}) with
$\kappa = 1$, whereas those for pressure fluctuations and for 
fluid particle accelerations with $\kappa = 2$.
The PDF for energy dissipation rates is given with $\kappa = 3$.

\section{Analysis of experiments}

\begin{figure}[thbp]
\begin{center}
\includegraphics[width=13cm]{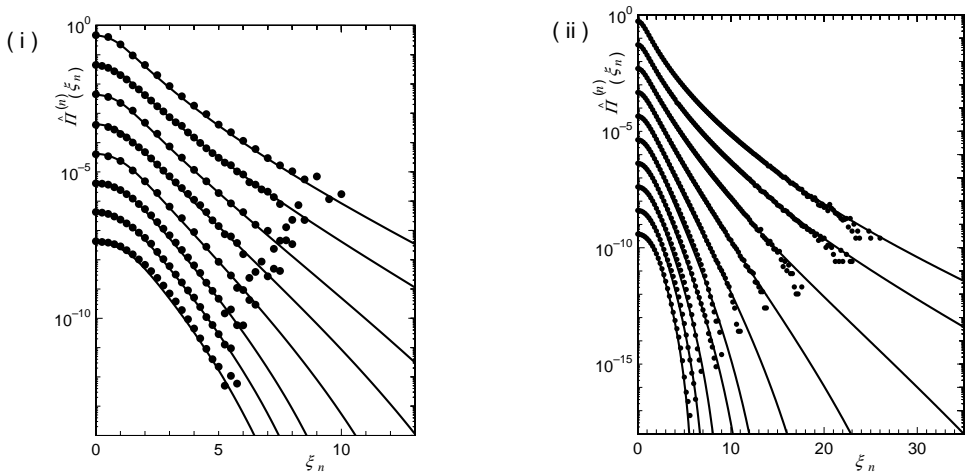}
\leavevmode
\caption{PDF's for {\it longitudinal} velocity fluctuations (closed circles;
symmetrized by taking averages of the left and the right hand sides data) 
observed by (i) Lewis and Swinney at $R_\lambda =262$ ($\rRe = 540\ 000$), and by
(ii) Gotoh et al.\ at $R_\lambda =381$.
For the experimental data, the distances $r/\eta = \ell_n/\eta$ 
are, from top to bottom: (i) 11.6, 23.1, 46.2, 92.5 208, 399, 830, 1440;
(ii) 2.38, 4.76, 9.52, 19.0, 38.1, 76.2, 152, 305, 609, 1220.
For the theoretical PDF's (solid line), from top to bottom:
(i) $\mu = 0.28$, $n\ (\bar{n}) =$ 
14 (10.7), 13 (9.7), 11 (8.7), 10 (7.7), 9.0 (6.6), 8.0 (5.6), 7.5 (4.6), 7.0 (3.8);
(ii) $\mu = 0.240$, $n\ (\bar{n}) =$ 
21.5 (15.8), 20.0 (14.3), 16.8 (11.1), 14.0 (8.31), 11.8 (6.11), 
10.1 (4.41), 9.30 (3.61), 8.10 (2.41), 7.00 (1.31), 6.00 (0.31).
For better visibility, each PDF is shifted by $-1$ unit along the vertical axis.
}
\label{velocity fluctuations log}
\end{center}
\end{figure}

The PDF's for velocity fluctuations measured by Lewis and Swinney \cite{Lewis-Swinney99} 
at $R_\lambda = 262$ for turbulent Couette-Taylor flow
and those extracted by Gotoh et al.\ from his DNS data \cite{Gotoh02} at $R_\lambda = 380$ 
are shown, respectively, in Fig.~\ref{velocity fluctuations log}-(i) and
in Fig.~\ref{velocity fluctuations log}-(ii).
They are analyzed with the derived formulae (\ref{PDF kappa large}) 
and (\ref{PDF kappa small}) with $\kappa = 1$ \cite{AA5,AA6}.

For Fig.~\ref{velocity fluctuations log}-(i), 
we adopted the reported value $\mu = 0.28$ \cite{Lewis-Swinney99} to calculate
the parameters $q = 0.471$, $\alpha_0 = 1.162$ and $X = 0.334$.
The dependence of $n$ on $r/\eta$ is extracted as \cite{AA5}
\be
n = -1.019 \times \log_2 r/\eta + 0.901 \times \log_2 \rRe
\label{n-roeta experimental}
\ee
with $\rRe = 540\ 000$. 
The Reynolds number is estimated with 
$
\ell_{\rm in} \approx 119.32 \mbox{ cm}
$
and the Kolmogorov scale 
$
\eta \approx 0.006 \mbox{ cm}
$. 
The energy input scale $\ell_{\rm in}$ is given by the size of 
experimental apparatus 
$
2 \pi \times 19.00 \mbox{ cm}
$
\cite{Lewis-Swinney99}.
The definition of the number of steps $\bar{n}$ 
within the energy cascade model is given by
$
\bar{n}= -\log_2 (r /\ell_{\rm in})
$
for the eddies whose diameter is equal to $r$.
By putting $r = \ell_n$, this gives us 
the relation between $\bar{n}$ and $n$ in the form
\be
\bar{n} = n - \log_2 (\ell_0/\ell_{\rm in}).
\label{n-nbar}
\ee

For Fig.~\ref{velocity fluctuations log}-(ii), we extracted the value $\mu = 0.240$ 
by analyzing the measured scaling exponents 
$\zeta_m$ of velocity structure function with the formula (\ref{zeta}),
which gives the values $q = 0.391$, $\alpha_0 = 1.138$ and $X = 0.285$.
Through the analyses of the PDF's for velocity fluctuations, we extracted
the formula for the dependence of $n$ on $r/\eta$: \cite{AA6,AA7}
\bea
n \aeq -1.050 \times \log_2 r/\eta + 16.74
\quad (\mbox{for } \ell_c \leq r),
\label{n-roeta L larger} \\
n \aeq -2.540 \times \log_2 r/\eta + 25.08
\quad (\mbox{for } r < \ell_c).
\label{n-roeta L less}
\eea
This shows that the inertial range is divided into two scaling regions separated by 
the characteristic length $\ell_{\rm c} /\eta = 48.26$ which is close to 
the Taylor microscale $\lambda /\eta = 38.33$ of the system.
The equation (\ref{n-roeta L larger}) is consistent with the picture of 
the energy cascade model in which each eddy breaks up into two pieces at 
every cascade steps, whereas (\ref{n-roeta L less}) indicates that, for $r < \ell_c$,
each eddy breaks up, effectively, into 1.33 \cite{AA7} pieces 
at every cascade steps.
This fact may be attributed to a manifestation of structural difference of eddies,
which can be checked by visualizing DNS eddies. Actually, one observes 
that DNS eddies with larger diameters than Taylor microscale $\lambda$ 
have rather round shapes, whereas eddies with smaller diameters 
compared with $\lambda$ have rather stretched shapes \cite{Tanahashi}.
The energy input scale for this DNS is estimated as the longest scale available
in the lattice with cyclic boundary condition, i.e.,
$\ell_{\rm in}/\eta = \pi/\eta \approx 1220$ with $\eta \approx 0.258 \times 10^{-2}$
\cite{Gotoh02}.
With this value of $\ell_{\rm in}$, (\ref{n-nbar}) with
(\ref{n-roeta L larger}) and (\ref{n-roeta L less}) gives the number of steps $\bar{n}$ 
within the energy cascade model for the DNS.

The analysis of the PDF for velocity derivatives reported 
by Gotoh et al.\ \cite{Gotoh02} are performed with $\mu = 0.240$.
We chose the value $n = 23.1$ ($\bar{n}= 17.4$). 
The corresponding length $r=\ell_n$ is calculated by (\ref{n-roeta L less}) to give 
$r / \eta = 1.716$.
This length may give us an estimate for the effective shortest length in 
processing the DNS data to extract velocity derivatives.
Note that it is about the same order of the mesh size 
$\Delta r /\eta = 2\pi/(1024 \times \eta) \approx 2.38$ \cite{Gotoh02}
of the DNS lattice.

\begin{figure}[thbq]
\begin{center}
\includegraphics[width=13cm]{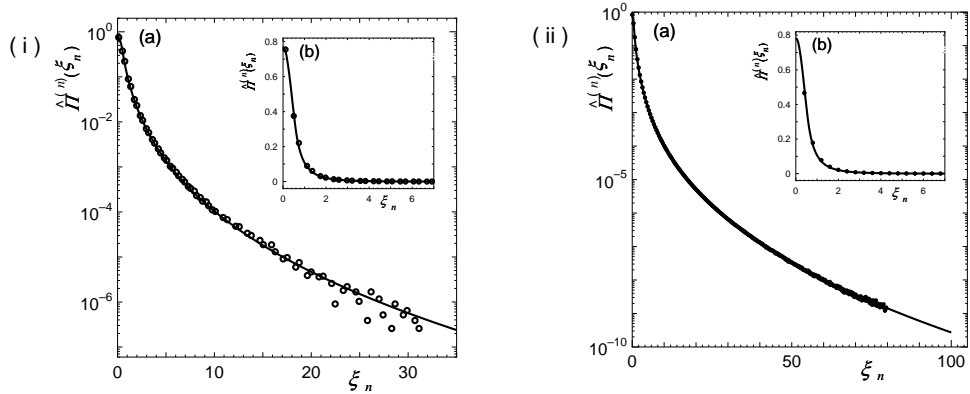}
\leavevmode
\caption{PDF's for fluid particle accelerations 
measured by (i)
Bodenschatz et al.\ at $R_\lambda =970$ (open circle),
and by (ii) Gotoh et al.\ at $R_\lambda = 380$ (closed circle),
plotted on (a) log and (b) linear scale.
The experimental data points both on the left- and right-hand sides 
are plotted altogether.
For the theoretical PDF's (solid line),
(i) $\mu = 0.250$ and $n=17.1$, and
(ii) $\mu = 0.240$ and $n=18.3$ ($\bar{n}=13.6$) (solid line).
\label{acceleration}}
\end{center}
\end{figure}

The PDF's for fluid particle accelerations 
measured by Bodenschatz et al.\ at $R_\lambda = 970$ \cite{EB01a,EB01b}
and those extracted out from the DNS data by Gotoh et al.\ 
at $R_\lambda = 380$ \cite{Gotoh02} 
are shown, respectively, in Fig.~\ref{acceleration}-(i) and
in Fig.~\ref{acceleration}-(ii), on log and linear scale \cite{AA9,AA10}.
They are analyzed with the derived formulae (\ref{PDF kappa large}) 
and (\ref{PDF kappa small}) with $\kappa = 2$ \cite{AA5,AA6}.

For Fig.~\ref{acceleration}-(i), we determined the value $n=17.1$ for this experiment
by substituting the reported value 7.1~cm for $\ell_0$ and 
the spatial resolution $0.5\ \mu\mbox{m}$ of the measurement for $\ell_n$
into its definition, 
$
n = \log_2 (\ell_0/\ell_n)
$.
The intermittency exponent $\mu = 0.250$ is extracted by the analysis of
the experimental PDF with the derived theoretical formula \cite{AA9,AA10}.
Then, we have 
the values of parameters: $q = 0.413$, $\alpha_0 = 1.144$ and $X = 0.297$.
The flatness of the PDF turns out to be 
$
F_{\mathrm{a}}^{(n)} \equiv \dbra \mathrm{a}_n^4 \dket
/\dbra \mathrm{a}_n^2 \dket^2
=\dbra \xi_n^4 \dket = 61.3
$
which is compatible with the value of the flatness $\sim 60$ 
reported in \cite{EB01a,EB01b}.

For Fig.~\ref{acceleration}-(ii), with $\mu = 0.240$ for this DNS,
we have the values of parameters: 
$q = 0.391$, $\alpha_0 = 1.138$, $X = 0.285$.
The analysis of the PDF obtained by the DNS gives the value 
$n = 18.3$ ($\bar{n}=12.6$).
Substitution of this value into (\ref{n-roeta L less}) gives 
the corresponding characteristic length $r/\eta = 6.36$ \cite{AA10}.
This may be the effective minimum resolution in cooking the DNS data to distill
accelerations.

\section{Discussions}

It was shown that the various experimental PDF's in turbulence
are analyzed successfully with the formulae (\ref{PDF kappa large}) 
and (\ref{PDF kappa small}) derived by the multifractal analysis.

From the above analyses, we see that the contribution 
of thermal fluctuation and/or measurement error to PDF's is restricted to 
smaller values, i.e., $\xi_n \leq \xi_n^*$.
In the case of the PDF's for velocity fluctuations, 
$
\xi_n^* = 1.10 \sim 1.32
$
($\alpha^* = 1.08$) 
for Lweis and Swinney \cite{Lewis-Swinney99},
and
$
\xi_n^* = 1.01 \sim 1.39
$
($\alpha^* = 1.07$)
for Gotoh et al.\ \cite{Gotoh02}.
As for the PDF for velocity derivatives by Gotoh et al.\ \cite{Gotoh02},
$
\xi_n^* = 0.982
$
($\alpha^* = 1.07$).
In the case of the PDF's for fluid particle accelerations,
$
\xi_n^* = 0.565
$
($\alpha^* = 1.01$)
for Bodenschatz et al.\ \cite{EB01a,EB01b},
and
$
\xi_n^* = 0.551
$
($\alpha^* = 1.005$)
for Gotoh et al.\ \cite{Gotoh02}.
Within the present approach, the intermittent large deviations
$\xi_n \geq \xi_n^*$ are a manifestation of the multifractal distribution of
singularities $\alpha$ due to the scale invariance 
of the Navier-Stokes equation for $\rRe \gg 1$.

There is no room to incorporate into the present multifractal analysis
the energy input scale $\ell_{\rm in}$ and the "system size" $\ell_0$.
The former is necessary to determine 
the number of steps $\bar{n}$ in the energy cascade model.
Once $\ell_{\rm in}$ is determined by investigating the structure of 
experimental apparatuses, the relation between $\bar{n}$ and 
the multifractal step $n$ is given by (\ref{n-nbar}).
Since main part of the multifractal analysis rests on the scale invariance,
the size of the system under consideration is assumed to be infinite, and 
therefore, the length $\ell_0$ may not have important physical meaning.
Actually, the empirical equation (\ref{n-roeta experimental}) extracted 
from the experimental PDF's for velocity fluctuations by Lewis and Swinney gives 
$
\ell_0 \approx 877 \mbox{ cm}
$
which is large compared with the largest size 
$\ell_{\rm in} \approx 119.32 \mbox{ cm}$
of the experimental apparatus.
For Gotoh's DNS, the empirical equation (\ref{n-roeta L larger}) gives 
$
\ell_0 /\eta \approx 63000
$
which is larger than the largest size 
$\ell_{\rm in} /\eta \approx 1220$ of the DNS lattice.
On the other hand, in the analysis of Bodenschatz's experiment, the assignment
of $\ell_0$ to the integral length scale 7.1~cm gives us reasonable value $n$.
It is worthwhile to note here that PDF's derived within 
the multifractal analysis seem to be sensitive to
the characteristic lengths such as the distance of two measuring points,
the space resolution in measurement and the mesh size of DNS.
How to put the information of characteristic lengths of experimental apparatus
into the multifractal analysis is one of the important future problems. 
It may be resolved when one succeeds to reveal the dynamical foundation
underlying the basis of the multifractal analysis starting an investigation by 
the Navier-Stokes equation with the energy input term.


The authors would like to thank Prof.~R.H.~Kraichnan and 
Prof.~C.~Tsallis for their fruitful and enlightening comments 
with encouragement, and are grateful to Prof.~E.~Bodenschatz and Prof.~T.~Gotoh
for the kindness to show their data prior to publication.

\end{document}